\documentclass{cernyrep}
\usepackage{texnames}

\usepackage{amsfonts,amssymb,textcomp}
\usepackage{cite}
\usepackage{graphicx}
\usepackage{adjustbox}

\usepackage{varwidth}
\usepackage{xcolor}

\begin{document}

\title{Introduction to Plasma Accelerators: the Basics}

\author{R.Bingham$^\text{1,2}$ and R. Trines$^\text{1}$}

\institute{$^\text{1}$Central Laser Facility, Rutherford Appleton Laboratory, Chilton, Didcot, Oxfordshire, UK\\
$^\text{2}$Physics Department, University of Strathclyde, Glasgow, UK}

\maketitle

\begin{abstract}
 In this article, we concentrate on the basic physics of
 relativistic plasma wave accelerators. The generation of relativistic
 plasma waves by intense lasers or electron beams in low-density
 plasmas is important in the quest for producing ultra-high
 acceleration gradients for accelerators. A number of methods are
 being pursued vigorously to achieve ultra-high acceleration gradi\-ents
 using various plasma wave drivers; these include wakefield
 acceler\-ators driven by photon, electron, and ion beams. We describe the basic equations and show how intense beams can generate
 a large-amplitude relativistic plasma wave capable of accelerating
 particles to high energies. We also demonstrate how these same
 relativistic electron waves can accelerate photons in plasmas.

\end{abstract}

%\submitto{\PPCF}
%\pacs{xxx, xxx}

\section{Introduction}
Particle accelerators have led to remarkable discoveries about the
nature of fundamental particles, pro\-viding the information that enabled
scientists to develop and test the Standard Model of particle
physics. The most recent milestone is the discovery of the Higgs boson
using the Large Hadron Collider---the $27\Ukm$ circumference $7\UTeV$
proton accelerator at CERN. On a different scale,
accelerators have many applications in science and technology,
material science, biology, medicine, including cancer therapy, fusion
research, and industry. These machines accelerate electrons or ions to
energies in the range of tens of megaelectronvolts to tens of gigaelectronvolts. Electron beams with
energies from several gigaelectronvolts to tens of gigaelectronvolts are used to generate intense
X-rays in either synchrotrons or free electron lasers, such as the
Linear Collider Light Source at Stanford or the European XFEL in Hamburg,
for a range of applications. Particle accelerators developed in the last
century are approaching the energy frontier. Today, at the terascale,
the machines needed are extremely large and costly; even the smaller-scale lower energy accelerators are not small. The size of a
conventional accelerator is set by the technology used to accelerate
the particle and the final energy required. In conventional
acceler\-ators, radio-frequency microwave cavities support the electric
fields responsible for accelerating charged particles. In these
accelerators, owing to electrical breakdown of the walls, the electric
field is limited to about $100\UMV\,\UmZ^{-1}$. For more than 30 years, plasma-based
particle accelerators driven by either lasers or particle beams have
shown great promise, primarily because of the extremely large accelerating
electric fields that they can support, about a thousand times greater than
conventional accelerators, leading to the possibility of compact
structures. These fields are supported by the collective motion of
plasma electrons, forming a space charge disturbance moving at a speed
slightly below $c$, the speed of light in a vacuum. This method of
particle acceleration is commonly known as plasma wakefield
acceleration.

Plasma-based accelerators are the brainchild of the late John Dawson
and his colleagues at the Uni\-versity of California, Los Angeles, and are being investigated worldwide with a great deal of
success. Will they be a serious competitor and displace the
conventional `dinosaur' variety? The impressive results that have so
far been achieved show considerable promise for future plasma
accelerators at the energy frontier, as well as providing much smaller
`table-top' ion and electron accelerators. Research on plasma-based
accelerators is based on the seminal work by the late John Dawson and his collaborator Toshi Tajima \cite{taj79}. The main advantage of a plasma-based accelerator is that it can support accelerating electric fields
many orders of magnitude greater than conventional devices that suffer
from breakdown of the waveguide structure, since the plasma is already
`broken down'. The collective electric field $E$ supported by the
plasma is determined by the electron density $E \propto n^{1/2}$,
where $n$ is the electron density, and is known as an electron plasma
wave; the collective electric fields are created by a drive beam that
may be either a laser or a charged particle beam. These electron plasma
waves travel with a phase speed close to the speed of the drive
beam. The electric field strength $E$ of the electron plasma wave is
approximately determined by the electron density, $E \propto n^{1/2}$,
where $n$ is the density in $\Ucm^{-3}$; for example, a plasma with
density $10^{18}\Ucm^{-3}$ can support a field of about $10^9\UV\,\UcmZ^{-1}$, a
thousand times greater than a radio-frequency accelerator. This translates to a
reduction in size of the accelerator and a reduction in cost.

The original plasma accelerator schemes investigated in the 1980s and
1990s were based on a long-pulse laser. Short-pulse lasers did not exist
because chirped pulse amplification had not yet been demonstrated in
the optical regime, only in the microwave regime. Experiments used the
beat-wave mechanism of Tajima and Dawson  \cite{taj79}, where two laser beams with a
frequency difference equal to the plasma frequency drive a
large-amplitude plasma wave. This changed when the process of chirped pulse amplification was ported from microwaves to laser
beams by Strickland and Mourou \cite{strickland1, strickland2}. Suddenly, laser
pulses could be produced that were shorter than the plasma wavelength
(or skin depth) $c/\omega_\textrm{p}$, where $\omega_\textrm{p}$ is the electron plasma
frequency. This led to a dramatic change in the shape of the
wakefield, from a `density ripple' with many periods to a one- or
two-period `bubble-shaped' wakefield \cite{pac91,rosenzweig}. The
regime where the pulse length of the driving laser or particle beam is
of the order of the plasma wavelength is commonly called the
`bubble' or `blowout' regime.  Most laser-driven and
particle-driven particle acceler\-ator experiments today are in this
regime, and are commonly known as laser wakefield or
beam-driven plasma wakefield acceler\-ators. In the laser wakefield
accelerator, the radiation pressure of a short, intense laser beam
pushes plasma electrons forward and aside, creating a positively
charged ion column. As the laser beam passes the displaced electrons
snap back, owing to the restoring force of the ions, and overshoot,
setting up a plasma density modulation behind the laser pulse. Similar
plasma wakefields are set up by relativistic charged particle beams
propagating through uniform plasma. A number of reviews on
electron acceleration by laser-driven or beam-driven plasma waves have
been published \cite{review1,review2,review3,review4,review5}.

Early experiments produced beams with large energy spread, but in 2004
three independent groups in three different countries demonstrated
laser wakefield acceleration producing mono-energetic electron beams
with good emittance using short-pulse lasers
\cite{dreambeam1,dreambeam2,dreambeam3}, a result predicted by Pukhov
and Meyer-ter-Vehn  \cite{pukhov1} and Tsung \textit{et al.} \cite{tsung04}.  Many groups
worldwide now routinely produce electron beams at gigaelectronvolt energies using
this scheme \cite{leemans06,kneip09,leemans14}. Similar plasma
wakefields are set up by relativistic charged particle beams
propagating through uniform plasma \cite{chen85,caldwell}. In 2007,
Joshi's group at the Uni\-versity of California demonstrated
acceleration of electrons in metre-long plasma columns using a SLAC
charged particle beam as a driver. This resulted in particles near the
back of the electron beam doubling their energy from $42\UGeV$ to $85\UGeV$
in a $1\Um$ long lithium plasma \cite{blum07}. This is a remarkable
result, since it takes $3\Ukm$ of the SLAC linac to accelerate electrons
to $42\UGeV$. The plasma beam-driven wakefield is incorporated into the
latest round of experiments at SLAC by a consortium called the
Facility for Advanced Accelerator Experimental Tests (FACET)
\cite{review3}. Both electron and positron beams are accelerated in
this facility \cite{litos14,corde}. Beam-driven plasma wakefields also
underpin the proton beam-driven wakefield experiment, AWAKE, which
will use the proton beam from CERN's Super Proton Synchrotron and a $10\Um$ long plasma column to produce gigaelectronvolt electrons
\cite{caldwell,assmann}.

Today most experiments are conducted in the bubble regime. This
includes experiments at the Berkeley Laboratory Laser Accelerator
Center at Lawrence Berkeley Laboratory \cite{bella,leemans14}
and the Ruther\-ford Appleton Laboratory's Central Laser Facility
\cite{kneip09,kneip,savert15}, as well as many other laser-plasma
acceler\-ator experiments around the globe. These experiments have
demonstrated mono-energetic electron beams at the gigaelectronvolt
scale and planned experiments using lasers will demonstrate
acceleration of electrons to $10\UGeV$. Recently, FACET experiments
demonstrated high efficiency in electron beam production, where the
energy transfer from the wakefield to the accelerated bunch exceeded
30\% with a low energy spread \cite{litos14}. Similar impressive
results using positrons have also been demonstrated \cite{corde}.

Despite the successes of these experiments, it is still necessary to
improve beam quality, in par\-ticu\-lar, to produce low energy spread and low
emittance, and improve beam focusing. Most of the experiments are guided by
plasma simulations that involve high-performance computing clusters. Com\-monly used
simulation codes include the particle-in-cell codes Osiris
\cite{osiris1,osiris2,osiris3}, VLPL \cite{vlpl}, Vorpal \cite{vorpal},
and Epoch \cite{epoch}. These simulations have already predicted that
between 10 and $50\UGeV$ electron beams can be created in one stage of a
plasma accelerator.

If plasma accelerators are to take over from conventional machines, a
great deal of effort still needs to be put into efficient
drivers. Suitable laser efficiencies and pulse rates seem likely
with diode-pumped lasers or with fibre lasers, but effort has to be
put into these schemes to meet the requirements necessary to drive a
wakefield. For beam-driven systems, electron beams at $100\UGeV$ and
proton beams with teraelectronvolt energies are required. These
exist at the Large Hadron Collider for protons and at FACET for electrons. For an
electron--positron system, a key challenge is positron acceleration;
some groups are investigating positron acceleration in
wakefields. Alternatively, an electron-proton collider or a
photon--photon ($\gamma$--$\gamma$) collider could be built, doing away
with the need for positrons, thus saving time and effort.

In the next section, we will discuss the short-pulse laser wakefield
accelerator scheme. We will present basic analytical theory that
lays the groundwork for all subsequent investigations into
laser-driven and beam-driven wakefield acceleration, and provide
results from particle-in-cell simulations of three-dimensional
laser wakefield acceleration.

\section{The laser wakefield accelerator (LWFA)}
In the laser wakefield accelerator (LWFA), a short laser pulse, whose
frequency is much greater than the plasma frequency, excites a wake of
plasma oscillations (at $\omega_{\textrm{p}}$), owing to the ponderomotive force,
much like the wake of a motorboat. Since the plasma wave is not
resonantly driven, as in the beat-wave, the plasma density does not have
to be of a high uniformity to produce large-amplitude waves. We start
from an intense laser pulse with electric field amplitude $E_0$ and
frequency $\omega_0$, and define $a_0 \equiv eE_0/(m_\textrm{e} \omega_0 c)$.
As this pulse propagates through an underdense plasma, $\omega_0 \gg
\omega_{\textrm{p}}$, the relativistic ponderomotive force associated with the
laser envelope, $F_{\textrm{pond}} \simeq - \frac{1}{2} m c^2 (\nabla
a_0^{2})/\sqrt{1+a_0^2}$, expels electrons from the region of the laser
pulse and excites electron plasma waves. These waves are generated as
a result of being displaced by the leading edge of the laser pulse.
If the laser pulse length, $c\tau_{\textrm{L}}$, is long compared with the
electron plasma wavelength, the energy in the plasma wave is
re-absorbed by the trailing part of the laser pulse.  However, if the
pulse length is approximately equal to or shorter than the plasma
wavelength $c\tau_{\textrm{L}} \simeq \lambda_{\textrm{p}}$, the ponderomotive force
excites plasma waves or wakefields, with a phase velocity equal to the
laser group velocity, which are not re-absorbed.  Thus, any pulse with a
sharp rise or a sharp fall on a scale of $c/\omega_{\textrm{p}}$ will excite a
wake. With the development of high-brightness lasers, the laser
wakefield concept first put forward by Tajima and Dawson \cite{taj79}
in 1979 has now become a reality.  The focal intensities of such
lasers are $\geq 10^{19}$\UW\,\UcmZ$^{-2}$, with $a_0 \geq 1$, which
is the strong non-linear relativistic regime. Any analysis must,
therefore, be in the strong non-linear relativistic regime and a
perturbation procedure is invalid.

The maximum wake electric field amplitude generated by a plane-polarized pulse has been given by Sprangle \textit{et al.} \cite{spr88} in the one-dimensional limit as $E_{\textrm{max}} = 0.38 {a_0^2 \left(1 +
  a_0^2/2 \right)^{-1/2}} \sqrt{n_0}\UV\,\UcmZ^{-1}$. For $a_0
\approx 4$ and $n_0 = 10^{18}\Ucm^{-1}$, then $E_{\textrm{max}} \approx 2\UV[G]\,\UcmZ^{-1}$, and the time to reach this amplitude level is of the order of
the laser pulse length.

\subsection{Model equations describing laser wakefield
excitation}
To understand the laser wakefield excitation mechanism, it is sufficient to use
a model based on one-fluid, cold relativistic hydrodynamics, and
Maxwell's equations, together with a `quasi-static' approxi\-mation, a set of two
coupled non-linear equations describing the self-consistent evolution in one dimen\-sion of
the laser pulse vector potential envelope, and the scalar potential of the
excited wakefield.  Starting from the equation for electron momentum,
\begin{equation}
{\partial{\bf p} \over {\partial t}} + v_{z} {\partial {\bf p} \over
{\partial z}} = - \left(e{\bf E} + {1 \over {c}} {\bf v} \times {\bf
  B}\right)~,
\label{eq:electronmomentum}
\end{equation}
where
\begin{equation*}
{\bf p} = m_{0}\gamma {\bf v}, \ \gamma = \left(1 +
p^{2}/m_{0}^{2}c^{2}\right)^{1/2},
\end{equation*}
$m_{0}$ and ${\bf v}$ being the electron rest mass and velocity.

In \Eref{eq:electronmomentum}, we have assumed that all quantities only depend on $z$ and
$t, z$ being the direction of propagation of the (external) pump and
\begin{equation}
{\bf E} = - {1 \over {c}} {\partial {\bf A}_{\perp} \over {\partial z}} -
\hat{z} {\partial \phi \over {\partial z}}\ ;\ {\bf B} = \nabla \times {\bf
A}_{\perp};\ \ {\bf A}_{\perp} = \hat{x}A_{x} + \hat{y}A_{y}~,
\label{eq:fieldequations}
\end{equation}
where ${\bf A}_{\perp}$ is the vector potential of the electromagnetic pulse
and $\phi$ is the ambipolar potential due to charge separation in the plasma.

Using Eqs. (\ref{eq:electronmomentum}) and (\ref{eq:fieldequations}), the perpendicular component of the electron momentum
is found to be
\begin{equation}
{p_{\perp} \over {m_{0}c}} = {e \over {m_{0}c^{2}}} {\bf A}_{\perp} \equiv
{\bf a}(z,t)~,
\end{equation}
and we can write
\begin{equation}
\gamma = \left[ 1 + \left({p_{\perp} \over {m_{0}c}}\right)^{2} + \left(
{p_{z} \over {m_{0}c}}\right)^{2}\right]^{1/2} \equiv
\gamma_{a}\gamma_{\parallel}~,
\end{equation}
where
\begin{equation}
\gamma_{a} = \left(1 + {\bf a}^{2}\right)^{1/2};\ \ \gamma_{\parallel} =
\left(1 - \beta^{2}\right)^{-1/2}~,
\end{equation}
and $\beta = v_{z}/c$.

The equations derived from this model are now the longitudinal component
of \Eref{eq:electronmomentum}, the equation of continuity, Poisson's equation, and the wave
equation for ${\bf a}(z,t)$, which are given by
\begin{align}
{1 \over {c}} {\partial \over {\partial t}}
\left(\gamma_{a}\sqrt{\gamma_{\parallel}^{2} - 1}\right) + {\partial \over
{\partial z}}\left(\gamma_{a}\gamma_{\parallel}\right) &= {\partial \phi \over
{\partial z}};\ \varphi \equiv {e\phi \over {m_{0}c^{2}}}~, \label{eq:longitudinalcomponent}\\
{1 \over {c}} {\partial n \over {\partial t}} + {\partial \over
{\partial z}} \left(n \sqrt{1-1/\gamma_{\parallel}^{2}} \right) &= 0~, \label{eq:continuity}\\
{\partial^{2}\varphi \over {\partial z^{2}}} &= {\omega_{\textrm{p}0}^{2} \over
{c^{2}}} \left({n \over {n_{0}}} - 1\right)~, \label{eq:Poisson}\\
c^{2} {\partial^{2}{\bf a} \over {\partial z^{2}}} - {\partial^{2}{\bf a}
\over {\partial t^{2}}} &= \omega_{\textrm{p}0}^{2} {n \over {n_{0}}} {{\bf a} \over
{\gamma_{a}\gamma_{\parallel}}}~.\label{waveequationazt}
\end{align}

Assuming a driving pulse of the form
\begin{equation}
{\bf a}(z,t) = \frac{1}{2} {\bf a}_{0} (\xi , \tau)\textrm{e}^{-\textrm{i}\theta} +
\textrm{c.c.},
\label{eq:drivingpulse}
\end{equation}
where $\theta = \omega_{0}t - k_{0}z, \omega_{0}$ and $k_{0}$ being the central
frequency and wavenumber, $\xi = z - v_{\textrm{g}}t$, $v_{\textrm{g}} =
\partial\omega_{0}/\partial k_{0}$ is the group velocity and $\tau$ is a slow
time-scale, such that
\begin{equation*}
a_{0}^{-1} {\partial^{2}a_{0} \over {\partial\tau^{2}}} \ll \omega_{0}^{2}~.
\end{equation*}
Accounting for changes in the pump due to the plasma reaction, the wave
equation becomes
\begin{multline}
\left[2 {\partial \over {\partial\tau}}\left(\textrm{i}\omega_{0}a_{0} + v_{\textrm{g}}
{\partial a_{0} \over {\partial\xi}}\right) + c^{2} \left(1 -
v_{\textrm{g}}^{2}/c^{2}\right) {\partial^{2}a_{0} \over {\partial\xi^{2}}} +
2\textrm{i}\omega_{0} \left({c^{2}k_{0} \over {\omega_{0}}} - v_{\textrm{g}}\right)
{\partial a_{0} \over {\partial\xi}}\right]\textrm{e}^{-\textrm{i}\theta} + \textrm{c.c.}\\
= \left[c^{2}k_{0}^{2} - \omega_{0}^{2} + {n \over {n_{0}}}
{\omega_{\textrm{p}0}^{2}>
{\gamma_{a}\gamma_{\parallel}}}\right] a_{0} \textrm{e}^{-\textrm{i}\theta} +
\textrm{c.c.}~,\label{eq:pumpchangeswaveequation}
\end{multline}
where $\omega_{\textrm{p}0}$ is the plasma frequency of the unperturbed plasma.
Equations (\ref{eq:longitudinalcomponent}), (\ref{eq:continuity}), (\ref{eq:Poisson}), and (\ref{eq:pumpchangeswaveequation}) form the basic set for this problem in the
`envelope approximation'.

In the weak-pump, weakly relativistic regime, the solution has the
structure of a wakefield growing inside the electromagnetic pulse and
oscillating behind the pulse, with the maximum amplitude being reached
inside the pulse.  Using the quasi-static approximation, the time
derivative can be neglected in the electron fluid equations, Eqs. (\ref{eq:longitudinalcomponent}) and (\ref{eq:continuity}), yielding the following constants:
\begin{align}
\gamma_{a} \left(\gamma_{\parallel} - \beta_{0} \sqrt{\gamma_{\parallel}^{2}
-1}\right) - \varphi& = 1~,\label{constantgamma} \\
n\left(\beta_{0}\gamma_{\parallel} - \sqrt{\gamma_{\parallel}^{2} -1}\right)&
= n_{0}\beta_{0}\gamma_{\parallel}~,\label{constantn}
\end{align}
where $\beta_{0} = v_{\textrm{g}}/c$. The constants of integration have been
chosen in such a way that
\begin{equation*}
n = n_{0},\ \gamma_{\parallel} = 1,\ \varphi = 0~,
\end{equation*}
for
\begin{equation}
\gamma_{a} = 1, \qquad \left(|a_{0}|^{2} = 0\right)~.
\end{equation}

Using Eqs.~(\ref{eq:drivingpulse}) and (\ref{constantn}), the general system, Eqs.~(\ref{eq:longitudinalcomponent})--(\ref{waveequationazt}), can be written as two
coupled equations, describing
the evolution of the laser pulse envelope ${\bf a}_{0}$ and the scalar
potential $\varphi$:
\begin{align}
{\partial^{2} \varphi \over {\partial\xi^{2}}} &= {\omega_{\textrm{p}0}^{2} \over
{c^{2}}} G~,\label{eq:scalarpotential}\\
2\textrm{i}\omega_{0} {\partial a_{0} \over {\partial\tau}}  + 2c\beta_{0}
{\partial^{2}a_{0} \over {\partial\tau\partial\xi}} + {c^{2}\omega_{\textrm{p}0}^{2}
\over {\omega_{0}^{2}}} {\partial^{2}a_{0} \over {\partial\xi^{2}}} &=
-\omega_{\textrm{p}0}^{2}Ha_{0}~,\label{eq:laserpulseenvelopeevolution}
\end{align}
where
\begin{equation*}
G = {\sqrt{\gamma_{\parallel}^{2} -1} \over
{\beta_{0}\gamma_{\parallel}-\sqrt{\gamma_{\parallel}^{2}-1}}}\ ,\ H = 1-
{\beta_{0} \over {\gamma_{a}\left(\beta_{0}\gamma_{\parallel} -
\sqrt{\gamma_{\parallel}^{2}-1}\right)}}~.
\end{equation*}

This set of non-linear equations, Eqs.~(\ref{eq:scalarpotential}) and (\ref{eq:laserpulseenvelopeevolution}), is obtained
using a quasi-static approximation, which yields two integrals of the
motion, given by Eqs.~(\ref{constantgamma}) and (\ref{constantn}). The model is valid for
electromagnetic pulses of arbitrary polarization and intensities
$|a_{0}|^{2} \geq 1$.

Equations (\ref{eq:scalarpotential}) and (\ref{eq:laserpulseenvelopeevolution}) can be solved numerically in the
stationary frame of the pulse.  Equation (\ref{eq:scalarpotential}), Poisson's equation for the
wakefield, is solved with the initial conditions $\varphi = 0$,
${\partial \varphi / \partial \xi} = 0$ by a simple
predictor--corrector method.  The envelope equation, \Eref{eq:laserpulseenvelopeevolution}, describing the
evolution of the laser pulse, is written as two coupled equations for
the real and imaginary parts of $a_{0}$ and solved implicitly.

\begin{figure}
\includegraphics[width=\textwidth, clip=true, trim=200 750 70 500]{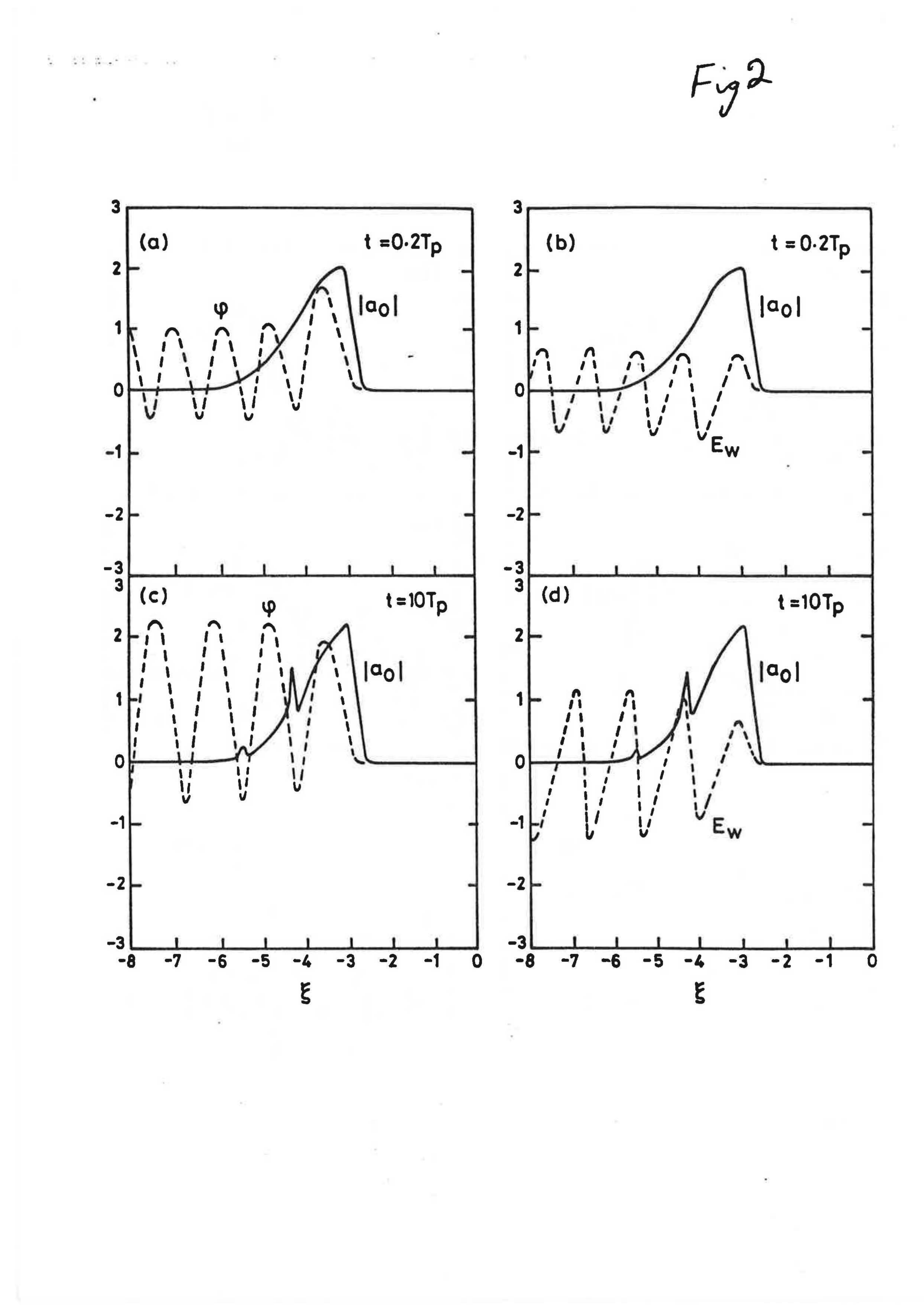}
\caption{The values of the magnitude of the normalized vector potential
$|a_0|$ (solid curves) and scalar potential $\varphi$ or
  wake electric field $E_\textrm{w}$ (dashed curves) versus position $\xi = z -
  v_\textrm{g} t$. Here, $|a_0^\mathrm{in} = 2|$, $\omega_{\mathrm{p}0}/\omega_0 = 1$,
  Gaussian rise $\sigma_\mathrm{r} = 0.25\lambda_\mathrm{p}$, Gaussian fall $\sigma_\mathrm{f }=
  1.5\lambda_\mathrm{p}$. Curves (a) and (b) are at time $t=0.2 T_\mathrm{p}$; (c) and
  (d) are at $t= 10 T_\mathrm{p}$.}
\label{fig:1}
\end{figure}

Numerical solutions of Eqs.~(\ref{eq:scalarpotential}) and (\ref{eq:laserpulseenvelopeevolution}) show the evolution of
the excited plasma wakefield poten\-tial $\varphi$ and electric field
$E_{\textrm{w}}$, as well as the envelope of the laser pulse $|a_0|$
\cite{bin92}, all in one spatial dimension. Solutions for a typical
laser and plasma configuration are shown in \Fref{fig:1}. There
is significant distortion of the trailing edge of the laser pulse,
resulting in photon spikes. The distortion occurs where the wake
potential has a minimum and the density has a maximum.  The spike
arises as a result of the photons interacting with the plasma density
inhomogeneity, with some photons being accelerated (decelerated) as
they propagate down (up) the density gradient. This effect was
predicted by John Dawson and his group, and is called the photon
accelerator \cite{wil89}. The distortion of the trailing edge
increases with increasing $\omega_{\textrm{p}0}/\omega_{0}$. The longitudinal
potential, $e\phi /(mc^{2}) > 1$ or $eE_{z} /(m_{\textrm{e}}\omega_{\textrm{p}0}c) > 1$,
is significantly greater than for fields obtained in the plasma
beat-wave accelerator. The field amplitude for the beat-wave
accelerator is limited by relativistic detuning, while no such
saturation exists in the laser wakefield accelerator.

\begin{figure}
\centering\includegraphics[width=0.7\textwidth]{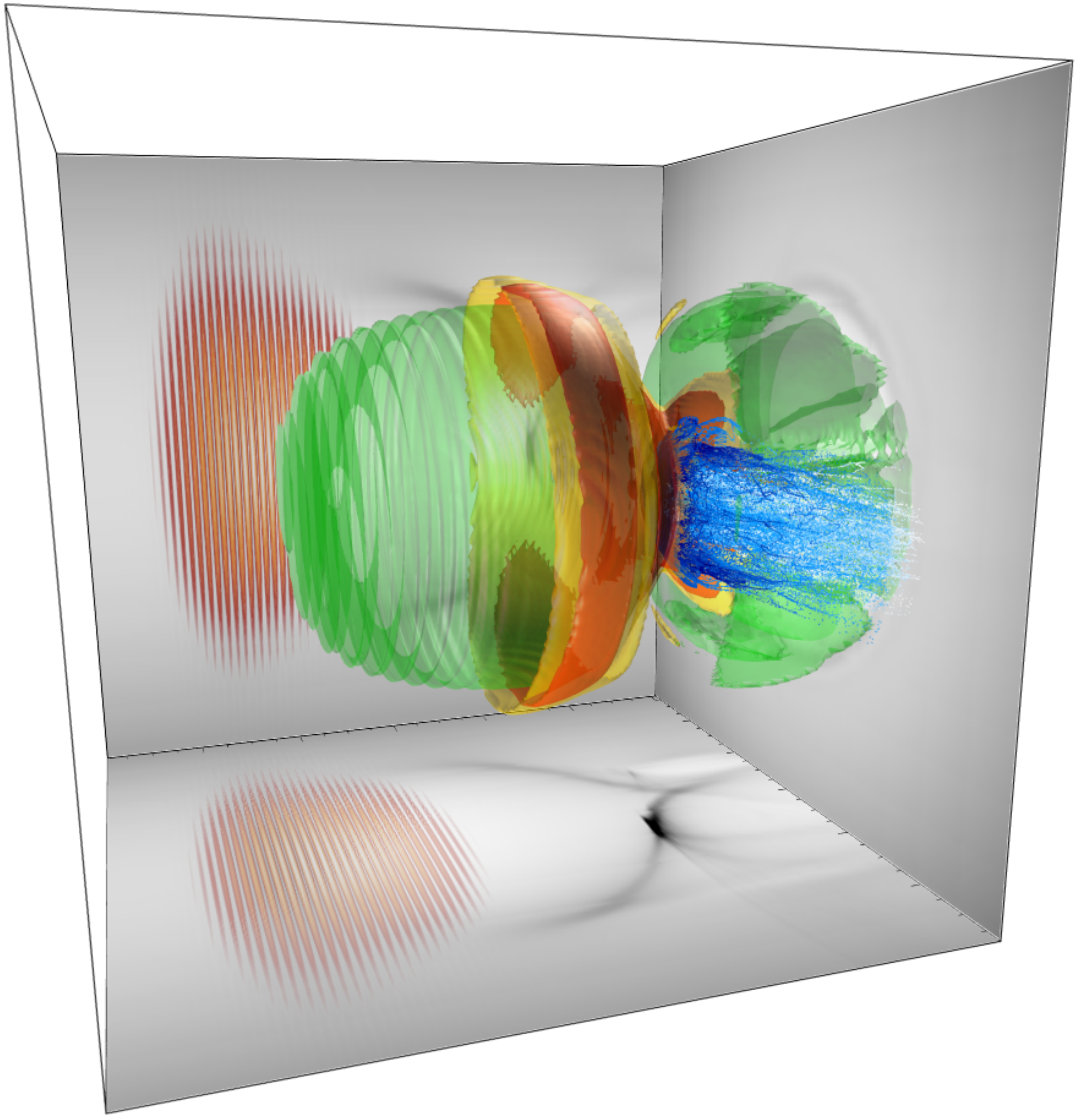}
\caption{An example of a three-dimensional wakefield generated by a
  short, intense laser pulse. Red and yellow, electromagnetic
  fields; green, electron density; blue, electrons trapped and
  accelerated by the wakefield. \textcopyright
 ~\ J.~Vieira, IST Lisbon, Portugal.}
\label{fig:2}
\end{figure}

When studying wakefields in more than one spatial dimension, the
transverse dimensions of the driving laser pulse or particle beam
become important, and the wakefield will assume a characteristic
`bubble' shape, provided the driver is sufficiently short. The bubble
regime of plasma-based acceleration has been studied extensively using
both analytical theory and full-scale numerical simulations
\cite{pukhov2,weilu1,weilu2,weilu3,weilu4,martins}. A typical example
of a three-dimensional laser-driven bubble-shaped wakefield can be
seen in \Fref{fig:2}. The electromagnetic fields are coloured
red and yellow and the background plasma electron density is coloured green,
while the population of electrons trapped and accelerated by
the wakefield is coloured blue.

\section{Photon acceleration}
Photon acceleration is the phenomenon whereby photons interacting with
a co-propagating plasma wave can increase or decrease their frequency,
and thus their energy. The group speed of photons in plasma is given
by $v_\textrm{g} = c^2 k/\omega = c\sqrt{1-\omega_\textrm{p}^2/\omega_0^2} < c$. Thus,
when the photon frequency increases, its group speed in plasma also
increases, hence the photon accelerates.  Photon acceleration
\cite{wil89,bingham_rsoc} is intimately related to short-pulse
amplification of plasma waves and is described by a similar set of
equations. It has been demonstrated that a relativistic plasma
wave can be generated by a series of short electromagnetic pulses
\cite{joh94}. These pulses have to be spaced in a precise manner to
give the plasma wave the optimum `kick'. Conversely, if the second or
subsequent photon pulses are put in a different position, 1.5 plasma
wavelengths behind the first pulse, the second pulse will produce a
wake that is $180^{\circ}$ out of phase with the wake produced by the
first pulse.  The superposition of the two wakes behind the second
pulse results in a lowering of the amplitude of the plasma wave.  In
fact, almost complete cancellation can take place.
The second laser pulse has absorbed some or all of the energy
stored in the wake created by the first pulse.  From conservation
of photon number density, the increase in energy of the second
pulse implies that its frequency has increased.  The energy in the
pulse is $W = N\hbar\omega$, where $N$ is the total number of
photons in the packet.  Wilks \textit{et al.} \cite{wil89} have shown that the
increase in frequency of the `accelerated' wave packet exactly
accounts for the loss in the energy of the accelerating plasma
wave, given by
\begin{equation}
{\delta\omega \over {\delta x}} = {\omega_{\textrm{pe}}^{2}k_{\textrm{p}} \over {2\omega}}
{\delta n \over {n_0}}~.
\end{equation}
Computer simulations by Wilks \textit{et al.} \cite{wil89} show that the laser pulse increased in frequency by
10\%  over a distance $237 c/\omega_{\textrm{p}}$.

Photon acceleration also plays a role in laser modulational
instability \cite{mod1,mod2,mod3}. This instability occurs when the
energy of a long laser pulse is bunched longitudinally by a
co-propagating plasma wave. The underlying process is periodic
acceleration and deceleration of the laser's photons by the density
fluctuations of the plasma wave. In turn, the laser's ponderomotive
force helps to enhance the plasma wave, creating a positive feedback
loop (see also Ref. \cite{pond1} for the relationship between the
ponderomotive force and the photon dispersion in plasma). The laser
amplitude modulation due to this instability can be seen in Fig.
\ref{fig:1}, frames (c) and (d).

Photon acceleration of laser light in laser wakefield interactions was
first observed by Murphy \textit{et al.} \cite{phacc1}. The role of photon
acceleration in the laser modulational instability was studied by
Trines \textit{et al.} \cite{phacc2,phacc3}, while aspects of the numerical
modelling of wave kinetics have been investigated by Reitsma \textit{et al.} \cite{phacc4,phacc5}.

Photon acceleration can be used as a single-shot diagnostic tool to
probe laser- or beam-driven wakefields and determine their
characteristics \cite{wil89,phacc1,phacc2}. This is currently pursued
within the framework of the AWAKE project \cite{assmann}. Promising
developments have been reported by Kasim \textit{et al.} \cite{kasim1,kasim2}.

\section{Relativistic self-focusing and optical guiding}

In the absence of optical guiding, the interaction length $L$ is
limited by diffraction to $L = \pi R$, where $R$ is the Rayleigh length
$R = \pi\sigma^{2}/\lambda_{0}$, and $\sigma$ is the focal spot
radius.  This limits the overall electron energy gain in a single
stage to $E_{\textrm{max}}\pi R$.  To increase the maximum electron energy gain
per stage, it is necessary to increase the interaction length.  Two
approaches to keeping high-energy laser beams collimated over a
longer region of plasma are being developed.  Relativistic
self-focusing can overcome diffraction and the laser pulse can be
optically guided by tailoring the plasma density profile, forming a
plasma channel.

Relativistic self-focusing uses the non-linear interaction of the
laser pulse and plasma results in an intensity-dependent refractive
index to overcome diffraction.  In regions where the laser intensity
is highest, the relativistic mass increase is greatest; this results in
a reduction of the fundamental frequency of the laser pulse.  The
reduction is proportional to the laser intensity.  Correspondingly, the
phase velocity of the laser pulse will decrease in regions of higher
intensity.  This has the effect of focusing a laser beam with a
radial Gaussian profile.  This results in the plane wavefront bending
and focusing to a smaller spot size.  Relativistic self-focusing has
a critical laser power threshold, $P_{\textrm{cr}}$, which must be exceeded; this is
given by \cite{sprangle-self-focus}
\begin{equation}
P \geq P_{\textrm{cr}} \sim 17 {\omega_0^{2} \over
  {\omega_{\textrm{p}}^{2}}}~\UW[G]~.
  \end{equation}
The laser must also have a pulse length $(\tau)$ that is shorter than
both a collision period and an ion plasma period, to avoid the
competing effects of thermal and ponderomotive self-focusing; this
condition is $\tau_{\textrm{L}} \leq 5\times10^{9} \sqrt{Z / n_0}\Ups$. The shape
of the self-focused pulse for intense short pulses is also
interesting; at the leading edge the non-linear response is not yet
established, there is a finite time for the electrons to respond, and
the front of the pulse propagates unchanged. The trailing edge of the
pulse compresses radially, owing to the non-linear relativistic
self-focusing.  It is only the trailing edge of the pulse that is
channelled.  Another way of forming a guided laser is to perform a
density cavity with one short-pulse laser forming a plasma density
channel.  Alternatively, the channel could be formed by a low-current
electron beam, which produces the same effect.

One drawback of the short-pulse laser--plasma interaction is
that, at the extreme intensities used, a hole in electron density can be
created.  This would prevent a wake from being created, or it might result in a very
small wake, if the residual density were many orders of magnitude
smaller than the original density.

\section{Discussion}

Plasma acceleration processes continue to be an area of active research. Initial studies of particle acceler\-ation have proved fruitful for current
drive schemes and laser accelerators.  Particle acceleration in strongly
turbulent plasmas is still in its infancy and requires a great deal more
research. This area of research is important in astrophysical and space
plasma.

Current and future experiments, however, are very far
from the parameter range of interest to high-energy physicists, who
require something like $10^{11}$ particles per pulse accelerated to
teraelectronvolt energies (for electrons), with a luminosity of $10^{-34}\Ucm^{-2}\,\UsZ^{-1}$ for acceptable event rates to be achieved. The
teraelectronvolt energy range is more than 100 times greater than a single
accelerating stage could provide at present; even if the interaction
length could be extended by laser channelling, there would still be
the requirement of multiple staging, and the need for more energetic lasers.

Researchers, realizing that the next collider will almost certainly be
a linear electron--positron collider, are proposing a novel way of
building such a device, known as the `plasma afterburner' concept
\cite{afterburner1,afterburner2}. Several groups are also developing
an entirely new type of electron lens, using focusing by a plasma, to
increase the luminosity of future linear colliders \cite{hir94}.

This plays on the fact that relativistic electron beams can be focused by a
plasma if the collisionless skin depth $c/\omega_{\textrm{pe}}$ is larger than the beam
radius.  Generally, when a relativistic electron beam enters a plasma, the
plasma electrons move to neutralize the charge in the beam on a $1/\omega_{\textrm{pe}}$
time-scale.  However, if the collisionless skin depth is larger than the beam
radius, the axial return current flows in the plasma on the outside of the
electron beam and the beam current is not fully neutralized, leading to the
generation of an azimuthal magnetic field.  Consequently, this self-generated
magnetic field pinches or focuses the beam in the radial direction.  This type
of lens exceeds conventional lenses by several orders of magnitude in focusing
gradient.

At present, laser-plasma accelerators are being used to accelerate
electrons from hundreds of megaelectronvolts to gigaelectronvolts, and to create intense
X-ray radiation via betatron motion of the accelerating electrons \cite{kneip}.
A future milestone to be achieved
will be the $10\UGeV$ energy level with good beam quality.
Electrons in this energy range are ideal as a driver for free
electron lasers; at the higher energies predicted, several
gigaelectronvolts, it is possible to produce an X-ray free electron laser capable of biological
investigations around the water window. Many other applications
of plasma-based acceleration are also already being discussed \cite{albert}.

%\section*{References}

\end{document}